\documentclass[twocolumn,amsmath,amssymb]{revtex4}
\usepackage{manfnt}

\usepackage{multirow}%
\usepackage{graphicx}% Include figure files
\usepackage{dcolumn}% Align table columns on decimal point
\usepackage{bm}% bold math

\begin{document}

\title{Negative thermal expansion near two structural quantum phase transitions}

\author{Connor A. Occhialini$^{1}$}
\author{Sahan U. Handunkanda$^{1,2}$}
\author{Ayman Said$^{3}$}
\author{Sudhir Trivedi$^{4}$}
\author{G. G. Guzm\'{a}n-Verri$^{5,6}$}
\author{Jason N. Hancock$^{1,2}$}
\affiliation{$^{1}$Department of Physics, University of Connecticut, Storrs, Connecticut 06269 USA}
\affiliation{$^{2}$Institute for Materials Science, University of Connecticut, Storrs, Connecticut 06269 USA}
\affiliation{$^{3}$Advanced Photon Source, Argonne National Laboratory, Argonne, Illinois 60349, USA}
\affiliation{$^{4}$Brimrose Technology Corporation, Sparks, MD 21152-9201, USA}
\affiliation{$^{5}$Centro de Investigaci\'{o}n en Ciencia e Ingenier\'{i}a de Materiales and Escuela de F\'{i}sica, Universidad de Costa Rica, San Jos\'{e}, Costa Rica 11501}
\affiliation{$^{6}$Materials Science Division, Argonne National Laboratory, Argonne, Illinois, USA 60439}

\date{\today}

\begin{abstract}
Recent experimental work has revealed that the unusually strong, isotropic structural negative thermal expansion in cubic perovskite ionic insulator ScF$_3$ occurs in excited states above a ground state tuned very near a structural quantum phase transition, posing a question of fundamental interest as to whether this special circumstance is related to the anomalous behavior. To test this hypothesis, we report an elastic and inelastic X-ray scattering study of a second system Hg$_2$I$_2$ also tuned near a structural quantum phase transition while retaining stoichiometric composition and high crystallinity. We find similar behavior and significant negative thermal expansion below 100K for dimensions along the body-centered-tetragonal $c$ axis, bolstering the connection between negative thermal expansion and zero temperature structural transitions. We identify the common traits between these systems and propose a set of materials design principles that can guide discovery of new materials exhibiting negative thermal expansion.
\end{abstract}

\pacs{PACS}

\maketitle

Negative thermal expansion (NTE) is an emerging area of material behavior discussed in chemistry, engineering, and physics which challenges conventional notions of lattice dynamics. Two routes to realizing NTE have been realized: one recipe is via a \textit{broadened phase transition} from some high-temperature phase to a higher-volume, lower temperature phase. Treatment with quenched disorder can smear the transition and achieve a gradual and tunable NTE evolution of lattice parameters. Examples of this type of NTE are found in InVar\cite{Guillaume1920}, anti-perovskites\cite{Takenaka2005,Qu2012a}, ruthenates\cite{Takenaka2017}, and charge-transfer insulators\cite{Azuma2011,Chen2013}. In contrast, a second type of NTE is realized from \textit{intrinsic dynamical origins}, also referred to as structural NTE (SNTE)\cite{Occhialini2017}, which is not obviously resultant from phase competition and does not require quenched disorder but seems to arise from intrinsic geometrical modes with tendencies to draw in the lattice dimensions when thermally activated. Unlike the broadened phase transition type, SNTE appears in a wide variety of lattice systems\cite{Dove2016,Mittal2018} without necessarily constraining the magnetic or electronic phase diagram. This freedom allows one to envisage new multifunctional materials with diverse mechanical, spin, orbital, thermal, electronic, superconducting, and more exotic order coexisting with NTE, potentially enabling the benefits of strain control to enable new types of order. 

ScF$_3$ is prominent among SNTE systems, forming in the so-called ``open'' perovskite (ReO$_3$-type) structure with a small, four-atom unit cell and cubic symmetry at all temperatures below the high melting point of 1800K (Fig 1a)\cite{Greve2010,Li2011,Handunkanda2015,Handunkanda2016,Hu2016,Piskunov2016,VanRoekeghem2016}. The linear coefficient of thermal expansion (CTE) of this material is isotropic, strongly negative, and persistent over 1000K temperature window\cite{Greve2010}. Combined computational and inelastic scattering work\cite{Li2011} has described the configurational potential for $R$ point distortions as having a nearly quartic form at small displacement, presenting an interesting limit of lattice dynamics. Further inelastic scattering investigations on single crystals aimed at exploring the consequences of this unusual situation discovered an incipient soft-mode instability\cite{Handunkanda2015} implying the development of a structural instability with a small extrapolated critical temperature $T_c$$<$0. This result implies that while similar compounds like TiF$_3$ realize a cubic-to-rhombohedral structural phase transition, the transition is never realized in ScF$_3$, except under an extremely small $<$1kbar hydrostatic pressure at cryogenic temperatures. %These observations are summarized in 
These special circumstances are contextualized in Figure 1a, which shows the global structural phase diagram of 3$d$ transition metal trifluorides BF$_3$ as a function of the B$^{+3}$ ionic radius, $r_B$. %which approaches -14ppm/K at low temperature. 
The occurrence of the endpoint of a structural phase boundary so near the ground state of a stoichiometric compound is extremely rare, as is the pronounced SNTE property of ScF$_3$. The confluence of these unusual circumstances raises the broad question of whether SNTE can arise as unusual behavior above structural quantum phase transitions (SQPTs) associated with transverse shifts of linking units between volume-defining vertices. To directly address this issue, we have explored the thermal expansion behavior and lattice fluctuation spectra in optical/detector quality single crystals of a second stoichiometric SQPT material Hg$_2$I$_2$, known colloquially as protiodide\cite{Amarasinghe2016}.

\begin{figure}
\begin{center}
\includegraphics[width=3.4in]{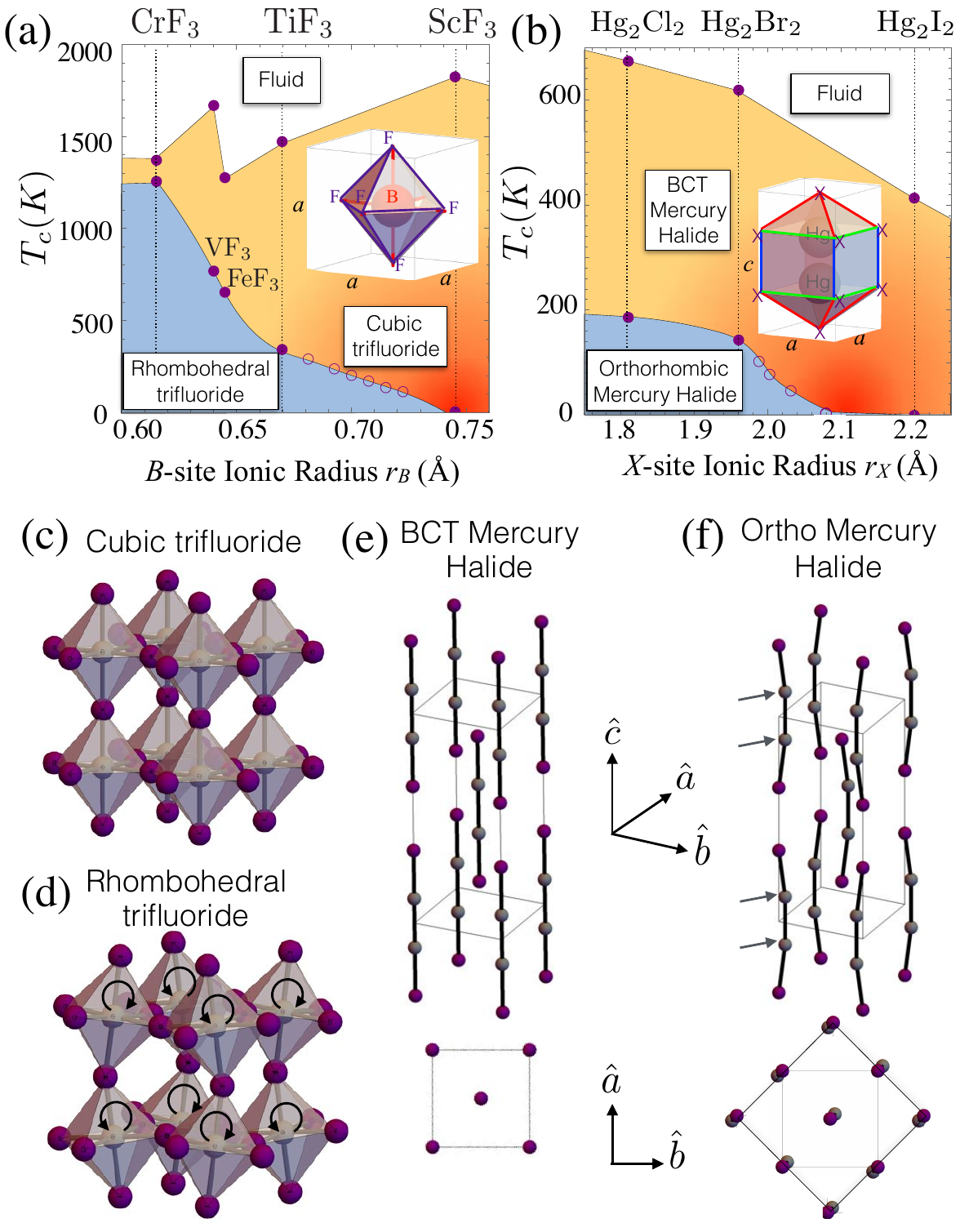}
\caption{Structural phase diagrams of (a) 3$d$ transition metal trifluorides BF$_3$ and (b) mercurous halides Hg$_2$X$_2$. Open circles represent solid solutions Sc$_{1-x}$Ti$_x$F$_3$\cite{Morelock2014} and Hg$_2$(Br$_{1-x}$I$_x$)$_2$\cite{Markov2007}, respectively. Insets show the basic volume-defining polyhedral units: (a) the BF$_6$ octahedron and (b) the elongated square dipyramid (ESD). (c-f) Schematic structures of the (c) cubic trifluoride (d) rhombohedral trifluoride (e) body-centered tetragonal mercury halide, and (f) orthorhombic mercury halide. The lower panels in (e) and (f) show views down the 001 axis, showing the shift pattern of the Hg dimer across the structural transition and gray line show the BCT unit cell and the black diamond in (f) shows the orthorhombic unit cell.}
%\label{ }
\end{center}
\end{figure}

Figure 1b summarizes the known structural phase diagram of the mercurous halides Hg$_2$X$_2$ (X=Cl,Br,I) as a function of the X$^-$ ionic radius\cite{Amarasinghe2016}, $r_X$. The high symmetry structure in this case is body-centered tetragonal (BCT; Fig 1e) and can be described as a dense packing of X-Hg-Hg-X linear molecules oriented along the tetragonal $c$ axis\cite{Havighurst1925}. The basic structural unit that defines the high-temperature unit cell is an elongated square dipyramid (ESD) formed as a cage with parts of 10 X$^-$ ions surrounding a Hg dimer oriented along $c$, shown in Figure 1b inset and Figure 4b. The structural transition to the orthorhombic phase can be described as a freezing of the Hg dimer transverse fluctuation in a staggered pattern (Fig. 1f) at the $X$ point of the BCT Brillouin zone.

\begin{figure}
\begin{center}
\includegraphics[width=3.4in]{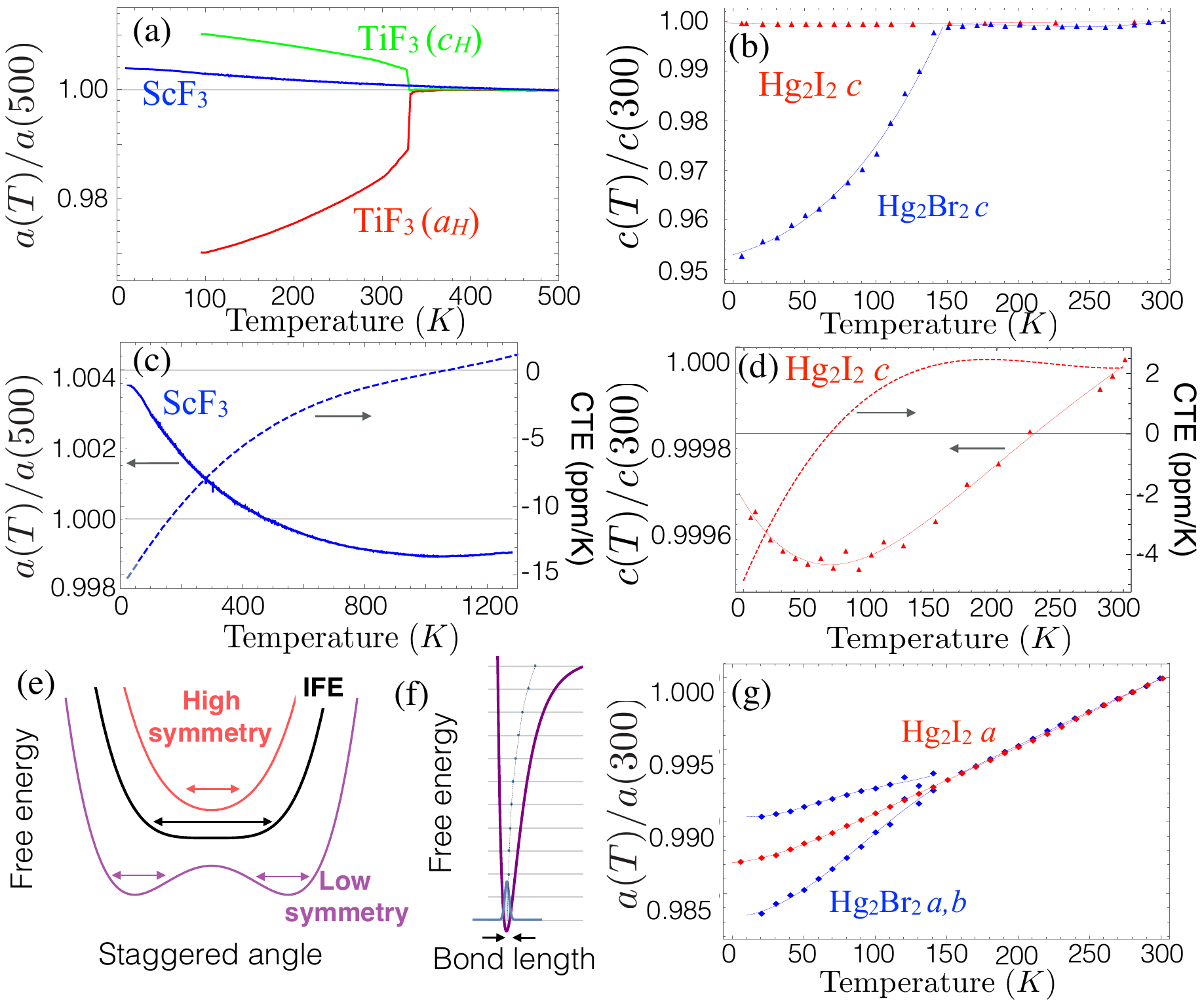}
\caption{Temperature-dependent lattice parameters of (a,c) ScF$_3$ and TiF$_3$ from references \cite{Greve2010,Handunkanda2015,Morelock2015}, and (b,d,f) Hg$_2$I$_2$, and Hg$_2$Br$_2$. The mercurous halide $c$ axis parameters were collected from the 004 and 008 reflections and the $a$,$b$ parameters are derived from the 300 and 030 twin reflections. The panel (e) shows a schematic free energy landscape of high, low, and transitioning systems near the SQPT. (f) Shows an intermolecular potential appropriate for bond-stretch coordinates. The mean separation is indicated by a dashed curve, illustrating the positive thermal expansion effect of this type of excitation. Horizontal arrows in (e) and (f) show the fluctuation domain in each case. (g) Planar lattice parameters for Hg$_2$Br$_2$ and Hg$_2$I$_2$.}
%\label{ }
\end{center}
\end{figure}

Single crystals of Hg$_2$I$_2$ and Hg$_2$Br$_2$ were prepared from purified materials using physical vapor deposition as previously reported\cite{Amarasinghe2016}. Diffraction and inelastic X-ray scattering (IXS) data were collected using the HERIX spectrometer in sector 30 of the Advanced Photon Source, Argonne National Laboratory. Figure 2 shows the lattice parameters of ScF$_3$\cite{Greve2010} and Hg$_2$I$_2$ along with data from their nearest realized structural transitions in TiF$_3$\cite{Morelock2014} and Hg$_2$Br$_2$, respectively, for comparison. In the case of TiF$_3$ (Fig 2a), the cubic-to-rhombohedral transition has profound effects on the lattice dimensions, showing a signature step of a first order transition and subsequent continuous order parameter development, shrinking $\sim$3\% between the structural transition temperature $T_c$ and base temperature\cite{Morelock2015}. The hexagonal $c$ axis on the other hand displays SNTE similar in magnitude to the negative CTE observed along the cubic $a$ axis of ScF$_3$ in the same temperature range. The lattice expansion of the realized transition in Hg$_2$Br$_2$ bears remarkable resemblance, with a significant reduction in tetragonal $c$ axis lattice parameter below the transition, changing 4.5\% between 150K and base temperature. The transition appears to be nearly continuous, consistent with prior work\cite{Boiko1992}. The $c$ axis of Hg$_2$I$_2$ however shows a strongly negative CTE below 100K shown in Figure 2d, reaching a significant low-temperature value of -5ppm/K. While negative values of the $c$-axis compressibility have been noted in mercurous halides\cite{Baughman1998}, SNTE in Hg$_2$I$_2$ has not been reported to our knowledge. 

The similarities to the fluorides are striking - both realized transitions TiF$_3$ and Hg$_2$Br$_2$ show a strong positive CTE below their respective transitions while the systems tuned near SQPTs show SNTE in high symmetry directions aligned with a linkage undergoing strong transverse fluctuations at the lowest temperatures. %A common structural motif between the two systems is the bridging subunit between aligned with dimensions that show SNTE: the bridging F ion along all three dimensions in ScF$_3$ and the bridging Hg dimer along the $c$ axis of the Hg$_2$I$_2$. 
Beyond the phase diagram and structural motifs, we note other gross similarities between these two material classes: they are ionic insulators with no reported magnetism or free carriers and retain high symmetry structures at all temperatures below their respective melting temperatures. Furthermore, both systems are driven to lower symmetry phases at modest pressure and ambient temperature (ScF$_3$ cubic-to-rhombohedral at $p_c$=7kbar and 300K\cite{Greve2010}; Hg$_2$I$_2$ body-center-tetragonal-to-orthorhombic at $p_c$=9kbar and 300K\cite{Markov2015}) or very low pressure at cryogenic temperatures (estimated $p_c$$\sim$1kbar for both systems at 4K\cite{Greve2010,Rehaber1982}). We note also analogous structural motifs: the bridging F ion in ScF$_3$, which occurs in all three spatial directions and the bridging Hg dimer, which is oriented along the $c$ axis in the mercurous halides.

Figs 2c,d show remarkable similarities in the functional form of the CTE of Hg$_2$I$_2$ and ScF$_3$, which strengthen significantly at low temperature, implying the relevant lattice modes lie at extremely low energy. Apart from exceptional cases\cite{Rechtsman2007,Kuzkin2014}, the influence of higher-energy bond-stretch excitations are expected to provide positive thermal expansion influences which generically compete with SNTE due to the short-range hardening and long-range softening of central force interionic potentials (Fig. 2f)\cite{Barrera2005,Takenaka2012}, which explains the overwhelming prevalence of positive thermal expansion in materials at ambient temperature. For Hg$_2$I$_2$, the magnitude of the $c$-CTE (-5ppm/K) is about one third of the maximum $a$-CTE in ScF$_3$ (-14ppm/K) and the range of thermal persistence is also reduced about 10-fold. This suppression can be partially explained by the much larger mass in the iodide case (ScF$_3$ density 2.53g/cm$^3$; Hg$_2$I$_2$ density 7.7g/cm$^3$), which significantly reduces the phonon bandwidth ($\sim$140meV in ScF$_3$\cite{Piskunov2016} and $\sim$25meV in Hg$_2$I$_2$\cite{Markov2015}) and activation energy for bond-stretch dynamics which compete with and ultimately overtake the SNTE effect. We point out further that quenched disorder is known to compete with SNTE\cite{Rodriguez2009}, and while we have restricted our attention to pure stoichiometric systems, the halide system is likely more disordered than the fluoride based on comparison of the rocking curve width (0.002$^\circ$ for ScF$_3$ and 0.13$^\circ$ for Hg$_2$I$_2$; see Supplemental Materials). A recent study of a SQPT achieved through chemical substitution shows prominent elastic and thermodynamic anomalies near the SQPT in solid solution LaCu$_{6-x}$Au$_{x}$\cite{Poudel2016} but does not report SNTE near the critical composition. In contrast, both ScF$_3$ and Hg$_2$I$_2$ lie very close to the critical endpoint of a structural phase boundary without the additional detrimental effects of quenched disorder. 

% and we note that even for pristine crystalline order, broad distributions of isotopes are inevitably present in naturally abundant Hg compounds, but less so in compounds that contain I, F, or Sc, which have nearly 100\% natural abundance of a single isotope. 

\begin{figure}
\begin{center}
\includegraphics[width=3.4in]{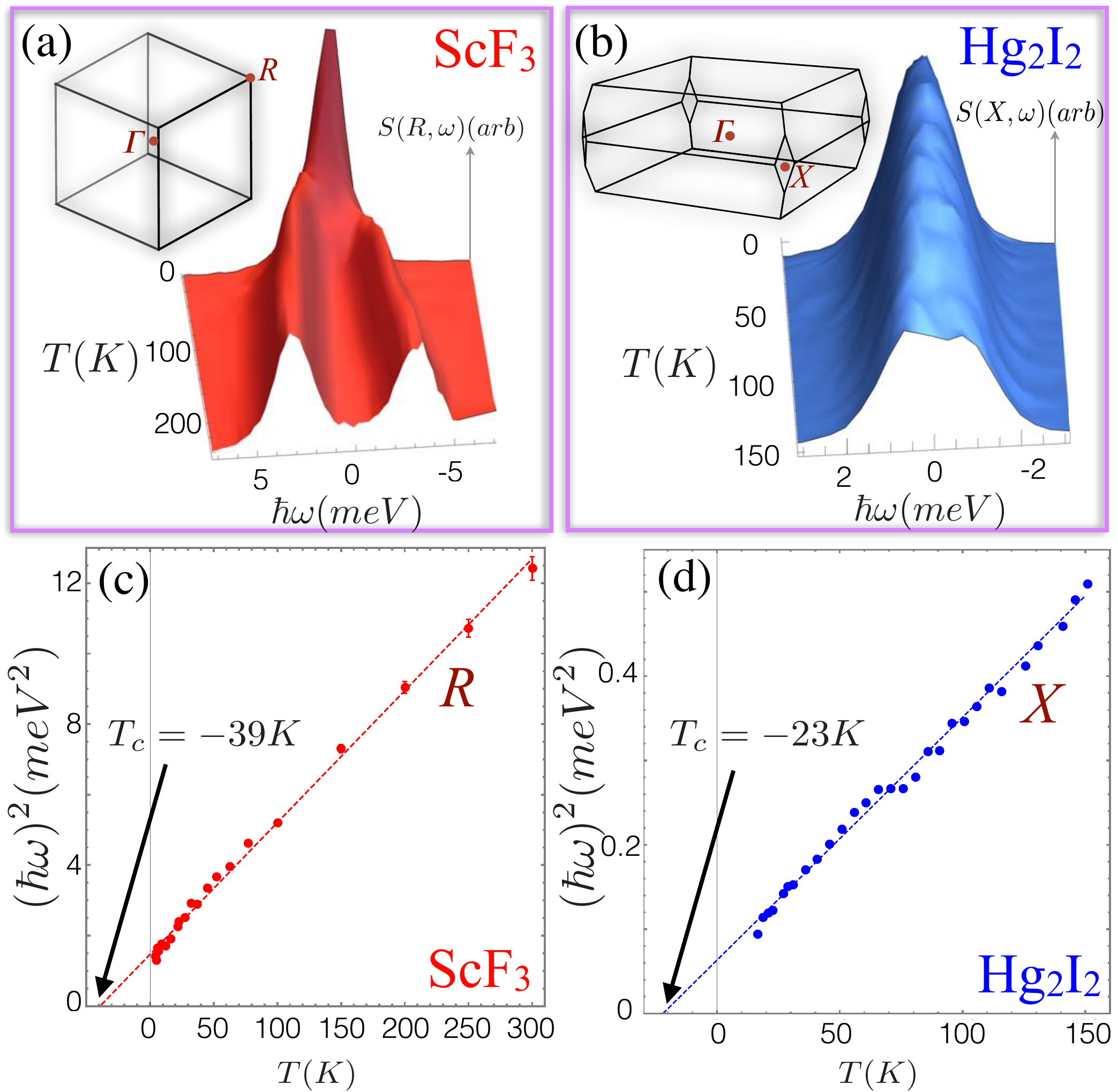}
\caption{(a) Dynamical structure factor for ScF$_3$ collected using inelastic X-ray scattering at the (2.5,3.5,0.5) reciprocal vector, corresponding to the $R$ point of the simple cubic Brillouin zone, shown as an inset. (b) Dynamical structure factor for Hg$_2$I$_2$ collected using inelastic X-ray scattering at the (2.5,3.5,0) reciprocal vector, corresponding to the $X$ point of the body-centered tetragonal Brillouin zone, shown as an inset. 
(c) Soft mode frequency squared determined from fits to the data in (a), showing the incipient ferroelastic state underlies the SNTE effect in ScF$_3$. (d) 
(d) Soft mode frequency squared determined from fits to the data in (b), showing the incipient ferroelastic state also underlies the SNTE effect in Hg$_2$I$_2$.}
\label{ }
\end{center}
\end{figure}

Figure 3 provides an experimental basis to demonstrate proximities to SQPTs in ScF$_3$ and Hg$_2$I$_2$, showing inelastic X-ray scattering spectra at the momentum points corresponding to the soft mode instabilities. The trifluoride low-temperature phase can be described as a staggered tilt of octahedra around the 111 axis (Figure 1f). This fluctuation has the ($\pi$,$\pi$,$\pi$) spatial texture of the $R$ point in the simple cubic Brillouin zone, shown in Figure 3a. Also shown is a surface plot of the dynamical structure factor $S(R,\omega)$ obtained using IXS\cite{Handunkanda2015}. At high temperature, a Stokes and anti-Stokes mode at low frequency of 3.4 meV softens considerably, approaching an extrapolated transition temperature $T_c$$\simeq$-39 K, as shown in Figure 3c. This singular point in the response function is suggestive of a flattening of the free energy landscape in an approach to an unrealized structural phase transition and strongly supports our identification of ScF$_3$ as near a SQPT. 

In the halide case, starting from the BCT phase, the relevant distortion to the low-temperature orthorhombic phase is a staggered shift of the Hg-Hg dimer from the ESD central axis in the 110 direction of the basal plane (Figure 1f) and the structural transition then corresponds to condensation of the transverse acoustic wave at the $X$ point of the BCT Brillouin zone\cite{Barta1975,Benoit1980,Boiko1992}. Figure 3b,d shows the evolution of $S(X,\omega)$ with temperature and an avoided condensation of a soft mode at the $X$ point of the body-centered tetragonal Brillouin zone in Hg$_2$I$_2$. $S(X,\omega)$ also shows Stokes/anti-Stokes mode pairs that imply a putative transition at $T_c$$\simeq$-23 K. This is fully consistent with previous studies on single crystals using energy-integrated diffuse X-ray scattering showing $T_c\sim$-20 K\cite{Markov2002}. We explore more detailed analysis of the influences driving these incipient structural transitions in each case below.

For the realized structural transitions in the mercurous halides X=Cl,Br\cite{Markov2002,Boiko1992} $T_c>$0 and the $T$=0 energy landscape consists of four minima corresponding to the possible saturated shifts of the Hg dimers (Fig 4d). For larger ionic radius X=I, these minima flatten and coalesce to the central axis, as no symmetry breaking is observed. 
The flattening of the energetic landscape however induces strong temperature-dependent fluctuations of the staggered shift, as is observed experimentally from X-ray diffraction data as large transverse dimensions of the Hg thermal ellipsoids, approaching $2\sqrt{U_{11}}$=0.38$\AA$ at $T$=150 K for Hg$_2$I$_2$\cite{Kars2012}. 

To gain further insight into the origin of the avoided transition in Hg$_2$I$_2$, we analyze trends among the bond lengths of the ESD, which is not regular, but rather compressed (Fig 1b inset), defined by X-X bond lengths of three varieties: the long X-X bonds (green) lying in the basal plane, the short X-X bonds (blue) oriented along the tetragonal $c$ axis, and apical X-X bonds (red). Figure 4b shows reported literature values of the lengths of the three types of X-X bond distance\cite{Havighurst1925,Havighurst1926,Hylleraas1927,Dorm1971,Calos1989,Kars2012}. The basal planar (green) X-X bonds cluster well\footnote{The tight clustering for the X-X ($a$) bonds is due to the simple relation to structure - this bond distance is simply the $a$ lattice parameter and can be determined from a single reflection and Bragg's law. The distances for the trifluoride show little scatter for the same reasons. The other internal distances require refinement of many allowed Bragg reflections over a larger angular range so show more scatter between reports.} with a clear trend far in excess of the diagonal dashed line that indicates simplest expectation based on sphere packing. The apical (red) bonds trend with ionic radius and stay near the X ionic diameter, indicating the apical half octahedron is flattened relative to a regular ESD but roughly satisfies simple packing conditions. On the contrary, the $c$-oriented (blue) X-X bond is near the ionic diameter only for X=Cl (calomel) and deviates significantly for X=Br and even more so for the SQPT material X=I, suggestive that the compression of the X-X bond plays a role in the energy flattening behind the SQPT and the NTE we report here. These short halide-halide bonds along $c$ are in a state of compression due to forces provided by the rest of the framework\footnote{This distance is much shorter than typical X-X distances obtained through searches of the international crystal database\cite{Belsky2002}.}. The natural source of this compression is a net tension in the X-Hg-Hg-X linear molecule along the $c$ axis through the ESD center. We propose that the origin of the coalescing energy minima can be viewed as an effect of two competing forces: the compressive stress on the $c$-oriented X-X bonds and the tensile stress on the X-Hg-Hg-X linear molecule by its environment. For X=I, these forces have a near-canceling effect which stabilizes the high-symmetry BCT structure with large transverse dimer fluctuations inside the ESD and the corresponding $c$-axis SNTE. Our identification of the compressive/tensile balance in Hg$_2$X$_2$ constitutes a context for the 1D ``tension effect"\cite{Schlesinger2008,Hu2016,Sanson2014,Dove2016}, which has been observed in cyanides\cite{Hibble2007,Hibble2010} in 1D and 2D. The Hg linking dimer is distinct from the linking CN complex in that it does not set up orientational order known to exist in a broad class of cyanides\cite{Hochli1990}.

Crossing the SQPT from the high symmetry side can be viewed as an increase in the number of zero modes as the potential landscape flattens in certain directions. In the language of structural mechanics, a change in zero mode count must be accompanied by a reduction in the number of states of self stress (SSS) in the ESD polyhedral unit\cite{Maxwell1864,Calladine1978,Kane2013}, a condition which increases its stability against deformation\cite{Fuller1975}. For the critical composition X=I, a critically-tensioned linear molecule inside the ESD unit exhibits large transverse fluctuations of the Hg dimer which exert tension on all bonds in the $c$ direction and leads to the observed negative thermal expansion along this axis which increases as the temperature is lowered (Fig. 2d). This size-induced stiffening is also apparent in the lowering of the melting temperature and triple points as the X ion grows, since the entropy of the BCT phase is reduced by the stiffening of the ESD, enabling the fluid phase to persist at lower temperature.

\begin{figure}
\begin{center}
\includegraphics[width=3.4in]{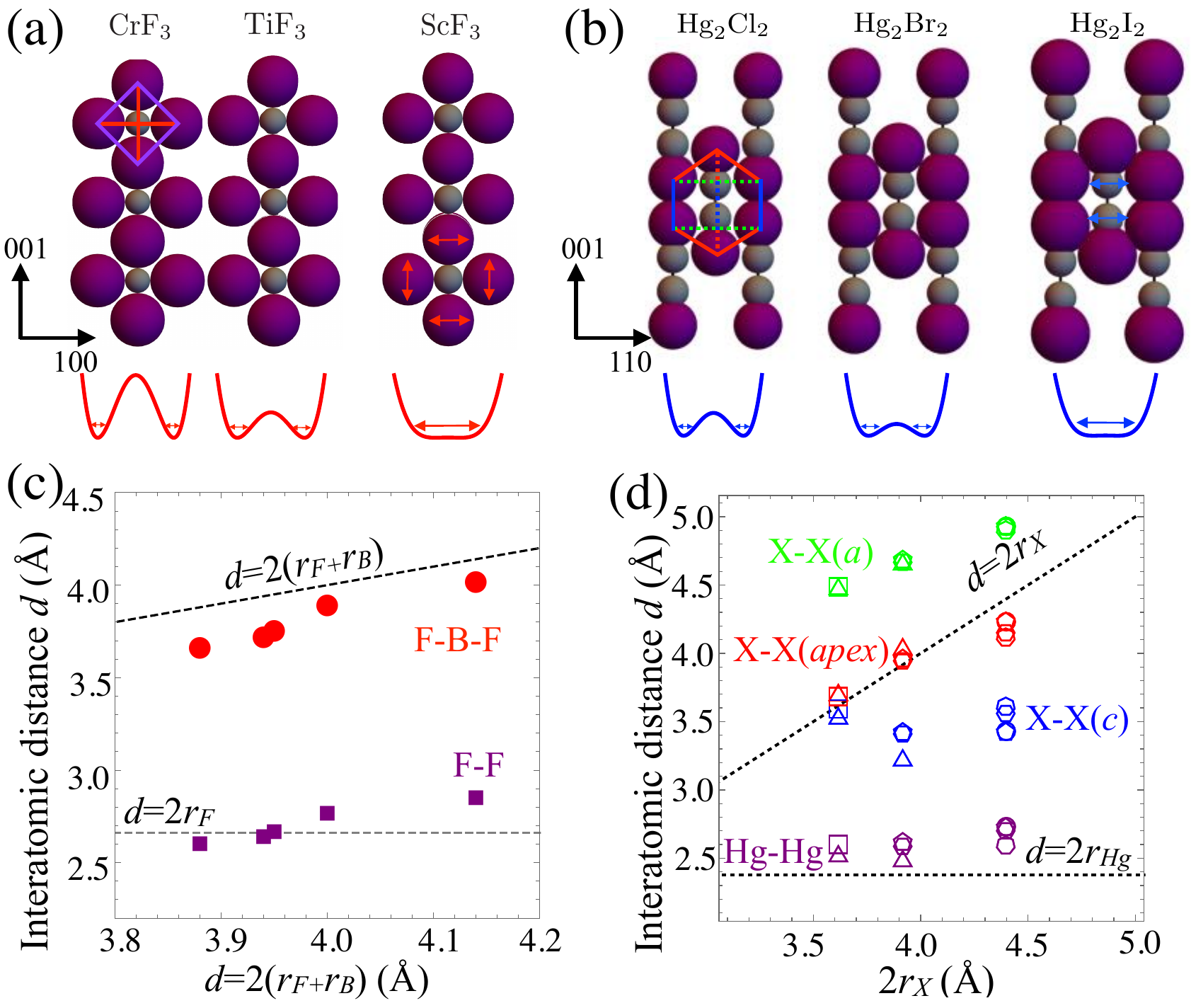}
\caption{(a) Development of the structures of the transition metal trifluorides, with purple spheres representing F$^-$ and silver spheres representing B$^{+3}$ ions. (b) Development of the structures of the mercurous halides with purple spheres representing X$^-$ and silver spheres representing Hg$^{-}$ ions. B$^{+3}$ or X$^-$ ionic radius increases from left to right in (a) and (b) respectively. Overlayed lines show the ESD of Figure 1b inset, where solid lines lie in the plane of the page and dashed lines show bonds that protrude out of the plane. (c,d) Comparison of the observed bond distances and expectations based on hard sphere packing (dashed lines) for (c) transition metal trifluorides in the cubic phase\cite{Daniel1990} and (d) mercurous halides in the BCT phase. Symbols in (d) are from corresponding references: triangles\cite{Dorm1971}, squares\cite{Calos1989}, pentagons\cite{Hylleraas1927}, hexagons\cite{Havighurst1926}, septagons\cite{Huggins1927}, and circles\cite{Kars2012}.}
\label{ }
\end{center}
\end{figure}

In a similar set of considerations, we analyze the evolution of the basic trifluoride octahedral subunit with $r_B$, used as a baseline for comparison of relative size effects. Figure 4a shows that to a good approximation, the unit cell dimension F-B-F ionic distance ($a$ lattice parameter) trends with $r_F$+$r_B$ while the nearest F-F distance grows to values significantly in excess of 2$r_F$. We note further that for $r_B$$>$0.75$\AA$ in the rare-earth class and beyond, trifluorides take on other crystal structures with B site coordination larger than $n$=6 in the orthorhombic tysonite ($n$=8), and hexagonal ($n$=9)\cite{Khitarov2001} structures. Together, these observations suggest an effective reduction of the overall stiffness of the BF$_6$ octahedron for large B ions through reduced interaction of the anions situated at the octahedral vertices. In this case, a large number of SSSs are removed as the octahedral unit loses integrity with increasing $r_B$.
Notably, fluctuations of the F$^-$ position transverse to the B-F-B bond are very large in ScF$_3$, approaching 2$\sqrt{U_{33}}$=0.24$\AA$\cite{Hu2016,Li2011,Piskunov2016} at $T$=150 K, consistent with this picture. In contrast to the mechanism in the mercurous iodide, we expect that the onset of pliancy of octahedral molecules stabilizes the cubic phase, as many states are available with the average cubic structure. This situation is also manifest in the high melting point of ScF$_3$, where the high entropy of transverse bond fluctuation competes with the fluid phase as high as 1800K. 

We note that incipient lattice instabilities and broadly systems tuned near SQPTs have recently attracted renewed interest\cite{Takahashi2017} in light of their use to develop a 50\% increase\cite{Stucky2016} in superconducting transition temperature of Nb-doped SrTiO$_3$ in an exceptional limit of the strong coupling theory\cite{Edge2015}, while the recent observation of electronic coupling to a substrate phonon in FeSe films raises questions regarding the possible role of substrate lattice fluctuations in stabilizing film superconductivity\cite{Lee2014a}. The common appearance of incipient soft modes also in SNTE materials potentially opens promising areas for future work combining SNTE and superconductivity to realize new emergent phases enabled by extremal strain conditions. Our results here suggest renewed importance of spectroscopic studies of known high-symmetry NTE materials such as ZrW$_2$O$_8$\cite{Hancock2004a,Bridges2014}, Zn(CN)$_2$\cite{Chapman2005a,Mittal2009}, and Ag$_3$[Co(CN)$_6$]\cite{Goodwin2008}.

The transition metal trifluoride and mercurous halide materials bear strong similarities besides the unusual strengthening of the SNTE effect at low temperature. In both cases, molecular units form a high-symmetry structure whose bonds are on average straight and are situated so as to define the linear dimensions of the crystal, $a$ in ScF$_3$ and $c$ in Hg$_2$I$_2$. On approach to zero temperature, high-energy bond-stretch excitations become frozen out quickly according to the Bose factor, while the soft mode angular fluctuations reduce more gradually due to the observed $\omega\propto\sqrt{T}$ dependence. These competing influences lead to trend of NTE strengthening at the lowest temperatures and conventional expansion at elevated temperatures and the soft mode is crucial to boost the weak NTE influence. From this point of view, ScF$_3$ has extremely strong and thermally-persistent NTE behavior due both to the stiff bond-stretch, three dimensional lattice system, and proximity to the SQPT. The incipient nature of the transition is vital to this condition to avoid a staggered strain symmetry breaking which disrupts the coupling of angle to dimension. We generalize this understanding in a proposal that candidate SNTE materials may be identified in systems which have ($i$) structural instability associated with a transverse linkage shift between-volume-defining vertices, ($ii$) soft modes and large fluctuations of the near the SQPT, ($iii$) low quenched disorder and stoichiometric composition, and ($iv$) relatively stiff bond-stretch excitations.

\begin{acknowledgments}
Work at the University of Connecticut was provided by National Science Foundation Award No. DMR-1506825 with additional support from the U.S. Department of Energy, Office of Science, Office of Basic Energy Sciences, under Award No. DE-SC0016481. C.A.O. acknowledges support from the Treibick family scholarship, managed by the Office of Undergraduate Research at the University of Connecticut. Work at the University of Costa Rica is supported by the Vice-rectory for Research under the project no. 816-B7-601 and work at Argonne National Laboratory is supported by the U.S. Department of Energy, Office of Basic Energy Sciences under contract no. DE-AC02-06CH11357.
\end{acknowledgments}

\bibliography{library}

\end{document}